\documentclass[sn-mathphys-num]{sn-jnl}
\pdfoutput=1
\usepackage{graphicx}%
\usepackage{multirow}%
\usepackage{amsmath,amssymb,amsfonts}%
\usepackage{amsthm}%
\usepackage{mathrsfs}%
\usepackage[title]{appendix}%
\usepackage{xcolor}%
\usepackage{textcomp}%
\usepackage{manyfoot}%
\usepackage{booktabs}%
\usepackage{listings}%
\usepackage{latexsym}
\usepackage{hyperref}
\usepackage{color}
\usepackage{bm}

\def\ba{\begin{eqnarray}}
\def\ea{\end{eqnarray}}

\begin{document}
 
\title[Cosmological evolution of matter with interacting dust fluids]{Cosmological evolution of matter with interacting dust fluids}
\author[1]{\fnm{Geoffrey} \sur{Okeng'o}}
\email{geffok@gmail.com}

\author*[2]{\fnm{Nceba} \sur{Mhlahlo}}\email{nmgaxamba@gmail.com}

\author[3]{\fnm{Roy} \sur{Maartens}}\email{roy.maartens@gmail.com}

\affil[1]{\orgdiv{Physics Department}, \orgname{University of Nairobi}, \orgaddress{\street{P. O. Box 30197-00100 GPO}, \city{Nairobi}, \country{Kenya}}}

\affil*[2]{\orgdiv{Physics Department}, \orgname{University of the Witwatersrand }, \orgaddress{\street{1 Jan Smuts Avenue}, \city{Johannesburg}, \postcode{2000}, \country{South Africa}}}

\affil[3]{\orgdiv{Department of Physics \& Astronomy}, \orgname{University of the Western Cape}, \orgaddress{\city{Cape Town}, \postcode{7535}, \country{South Africa}}}

\abstract{We split the total matter fluid into a bound (halo) component and an unbound (free particles) fluid component that is accreted by the halos. We adopt a different framework that treats the structure formation problem as a gravitational interaction between these virialised cold dark matter halos and the unbound inter-halo cold dark matter (and cold baryon) particles. This interaction involves in general an exchange of energy and momentum during the accretion process. We then explore the evolution of the average matter density and of large-scale structure formation, using a simplified phenomenological model that is based on results from extended Press-Schechter and N-body simulations. At high redshifts most matter is in diffuse form and is not part of the halos. As particles are accreted by the virialised halos, the particle number density decreases and that of the bound matter increases. We also present a general analysis of the background and linear perturbations for the  interacting fluids, showing in detail the energy and momentum exchange terms.}

\maketitle

\section{Introduction}

In the standard $\Lambda$CDM cosmological model, at late times when radiation may be neglected, the observed large-scale structure in the universe is understood to be based on cold dark matter (CDM) halos which  act as gravitational sinks for baryonic matter, leading to the formation of stars, galaxies and higher-level clustering of luminous matter. Here we  consider a simplified matter model in which
the
CDM is an elementary particle (e.g. the neutralino) and the baryons are cold. We neglect the complexities associated with non-cold baryonic matter by confining our analysis to linear scales. 
In this scenario, the first collapsed bound {(virialised)} objects are CDM halos of mass determined by the mass of the CDM particle -- e.g., approximately Earth mass for a neutralino particle
\cite{Angulo:2009hf}. Subsequently, halos grow by accretion of the unbound inter-halo particles. Moving forward in time, an increasing fraction of the total matter is captured in bound structures, with a spectrum of masses
\cite{Angulo:2009hf,Chan:2012jj,vanDaalen:2015msa,Shattow:2015ama,Angulo:2016qof,Zavala:2019gpq}.

Analytical models of structure formation are based on a Friedmann-Lem\^aitre-Robertson-Walker (FLRW) background with perturbations that grow under gravitational instability. The implicit assumption involved here is that the evolving mixture  of particles of different (and changing) masses may be treated as  a single pressure-free `dust' model, all of whose particles are considered as test particles -- i.e. effectively massless particles. Standard perturbation theory does not specify the masses of the dust particles and their changing {mass} spectrum. The agreement between the results of the perturbative analysis and those of N-body simulations on large enough scales suggests that this assumption is reasonable. Nevertheless, it is {interesting to} ask what the nature of the evolving spectrum of masses is and whether and how it affects cosmic evolution and structure formation. This question can be answered by N-body simulations \cite{Angulo:2009hf,Chan:2012jj,vanDaalen:2015msa,Shattow:2015ama,Angulo:2016qof,Zavala:2019gpq} but an analytical approach is also of interest. In this paper we adopt a simplified analytical approach to tackle this question (see e.g. \cite{Biagetti:2014afa,Biagetti:2015hva,Matsubara:2019tyb} for more sophisticated approaches).

Studies of interacting fluids (see e.g \cite{Valiviita:2008iv} for the theoretical foundations and \cite{Yang:2022csz,Wang:2024vmw} for recent work) have so far mainly focused on the ``dark sector'' interaction between dark matter and dark energy and have provided useful insights into the coincidence problem and structure formation, within general relativity.\footnote{Note that interacting fluid models can be considered as GR with a modified matter Lagrangian \cite{Skordis:2015yra,Akarsu:2023lre}.} 
A major difficulty confronting these models is the absence of a physically well-motivated non-gravitational interaction.
The interaction that we propose here is not intended in any sense as non-gravitational energy-momentum exchange -- on the contrary, it is the purely gravitational interaction of accretion.

In  other words, we treat the structure formation problem as a gravitational interaction between virialised CDM halos  and the unbound inter-halo CDM (and cold baryon) particles. This interaction is simply an exchange of energy and momentum during the accretion process.
However, given the nonlinear complexities of accretion in structure formation, we use a simple phenomenological model that is  based on an extended  Press-Schechter analysis,  tested with N-body simulations. To this end,  
we use a simple fit to the  results from \cite{Angulo:2009hf}. These results indicate that by redshift zero, $\sim 80-95\%$ of matter is within halos of all possible masses -- with $\sim 60-70\%$ of matter in halos of mass $\gtrsim 3\times 10^9\,M_\odot$ (see  Fig. 3 in \cite{Angulo:2009hf}). The percentages depend on the details assumed in  their models. These details are not important for our purposes, since our focus is on a simple analytical model that reflects the key qualitative features.

\section{The 2-fluid model}\label{equations}

The two  fluids are the halo fluid $A=h$ and the free (unbound) particle fluid $A=f$, each with a dust equation of state, $w_A=0$, and vanishing speed of sound  $c_{sA}=0$. The total matter density is
\ba \label{rmhf}
\rho_m=\rho_h+\rho_f\,,
\ea
which satisfies the background 
and perturbation equations of the standard $\Lambda$CDM model. For example, in the background
\ba \label{frw}
\mathcal{H}^2=\frac{8\pi G}{3}\,\bar{\rho}_m a^2+ \frac{\Lambda}{3}\,a^2\,,\quad {\bar\rho}'_m + 3\mathcal{H} \bar\rho_m=0\,,
\ea
where the prime denotes a conformal time derivative and $\mathcal{H}=a'/a$ is the conformal Hubble rate.

We assume a simplified model for the  density transfer from free particles to halos:
\begin{equation}
\frac{\rho_{h}}{\rho_{m}}=\mathcal{F}(\rho_{m}) \,. \label{eq:Frhom}
\end{equation} 
In other words, the fraction of mass in halos is given by a function of the total density. 

\subsection{Background equations}

In the background,
\ba
\bar{\rho}'_{h}+3\mathcal{H} \bar\rho_h &= & \bar\rho_m \bar{\mathcal{F}}'\,, \label{brhp}\\
\bar{\rho}'_{f}+3\mathcal{H} \bar\rho_f &= &- \bar\rho_m \bar{\mathcal{F}}' \,, \label{brfp}
\ea
where we used \eqref{rmhf} and \eqref{frw}. Note that $\bar{\mathcal{F}}'=-3\mathcal{H} \bar\rho_m \partial \bar{\mathcal{F}}/ \partial \bar\rho_m$.

The general background equations for two interacting fluids (see e.g. the extensive review in \cite{Valiviita:2008iv}) reduce in our case to 
\begin{eqnarray}
\bar{\rho}_{h}^{\prime}+3\mathcal{H}\bar{\rho}_{h}&=& a\bar{Q}_{h} \,, \label{eq:rhoh-prime}\\
\bar{\rho}_{f}^{\prime}+3\mathcal{H}\bar{\rho}_{f}&=& a\bar{Q}_{f}=-a\bar{Q}_{h} \,. \label{eq:rhof-prime}
\end{eqnarray} 
Here $\bar{Q}_A$ are the rates of energy density transfer: 
\ba \label{barq}
a\bar{Q}_h= \bar\rho_m \bar{\mathcal{F}}'=-a\bar{Q}_{f} \,.
\ea
 Clearly $\bar{Q}_h $ is positive since the halo fluid is accreting the free-particle fluid.
We can also rewrite these equations in terms of the dimensionless density parameters:
\begin{eqnarray}
\Omega_{h}^{\prime}&=&\frac{\mathcal{\bar{F}}^{\prime}}{\mathcal{\bar{F}}}\Omega_{h}+3\mathcal{H}\left(\Omega_{h}+\Omega_{f}-1\right)\Omega_{h} \,, \label{eq:omegah-prime}\\
\Omega_{f}^{\prime}&=&-\frac{\mathcal{\bar{F}}^{\prime}}{\mathcal{\bar{F}}}\Omega_{h}+3\mathcal{H}\left(\Omega_{h}+\Omega_{f}-1\right)\Omega_{f} \,. \label{eq:omegaf-prime}
\end{eqnarray}

\subsection{Linear perturbations}

Linear perturbation of \eqref{eq:Frhom} gives
\begin{eqnarray} \label{eq:deltah-deltam}
\delta_{h}-\delta_{m}&=&\frac{\delta \mathcal{F}}{\bar{\mathcal{F}}} 
     =\frac{\partial \ln \bar{\mathcal{F}}}{\partial \ln\bar{\rho}_{m}}\,\delta_{m} = \bar{\mathcal{G}}\,\delta_{m}
     \quad \text{where}\quad \bar{\mathcal{G}}\equiv\frac {\partial \ln \bar{\mathcal{F}}}{\partial \ln\bar{\rho}_{m}} \,.
\end{eqnarray} 
From this we obtain
\ba
\delta_{h} &=& \left(1+\bar{\mathcal{G}}\right)\delta_{m} \,, \label{eq:deltah1}
\\
\delta_{f}&=& \left(1-\frac{\Omega_{h}}{\Omega_{f}}\,\bar{\mathcal{G}}\right)\delta_{m} \,. \label{eq:deltaf1}
\ea
It follows from these equations that the power spectra of the halo fluid and the free-particle fluid are
\ba
P_{h}(a,k) &=& \left[1+\bar{\mathcal{G}}(a)\right]^2 P_{m}(a,k) \,, \label{powh}
\\
P_{f}(a,k)&=& \left[1-\frac{\Omega_{h}(a)}{\Omega_{f}(a)}\,\bar{\mathcal{G}}(a)\right]^2 P_{m}(a,k) \,, \label{powf}
\ea
where $P_m$ is the matter power spectrum.

Equations \eqref{eq:omegah-prime}--\eqref{powf} allow us to evade the solving of the perturbed energy-momentum conservation equations. Nevertheless, it is instructive to see how our equations correspond to the general perturbation equations  for coupled fluids.
In order to make this comparison, we choose the Newtonian gauge.
The perturbed  line element  at late times is then
\begin{equation}
{\rm d}s^{2}=a^{2}\left[-(1+2\Phi){\rm d}\eta^{2}+(1-2\Phi){\rm d}\bm{x}^2\right] \,, \label{eq:perturbed-line-elem}
\end{equation} 
where $\Phi$ is the gravitational potential.
The four velocity of fluid $A$ is
\begin{equation}
u^{\mu}_{A}=\frac{1}{a}\left(1-\Phi,\, \partial^{i}\upsilon_{A}\right),
\label{eq:u-mu}
\end{equation} 
where $\upsilon_{A}$ is the peculiar velocity potential, which vanishes in the background. 
The general perturbation equations for  two interacting fluids are based on \cite{Valiviita:2008iv} 
\begin{equation}
\nabla_{\nu}T^{\mu\nu}_{A}=Q^{\mu}_{A}\,,  \label{eq:nabla-T-mu-nu-A}
\end{equation} 
where $Q^{\mu}_{A}$ is the energy-momentum transfer four-vector, which satisfies $\sum_{A}Q^{\mu}_{A}=0$ from conservation of total energy-momentum.
In general the four-vector $Q^{\mu}_{A}$ has the form
\begin{equation}
Q^{\mu}_{A}=Q_{A}u^{\mu}+F^{\mu}_{A}\quad \text{where}\quad Q_{A}=\bar{Q}_{A}+\delta Q_{A}~~\text{and}~~ u_{\mu}F^{\mu}_{A}=0 \,. \label{eq:Q-mu-A}
\end{equation} 
Here $F^{\mu}_{A}$ is the momentum density transfer rate relative to the total four-velocity $u^{\mu}=a^{-1}(1-\Phi,\, \partial^{i}\upsilon)$ and $f_{A}$ is the intrinsic momentum transfer potential. It follows that
\begin{eqnarray}
Q^{A}_{0}&=&-a\,[\bar{Q}_{A}(1+\Phi)+\delta Q_{A}], \label{eq:Q-A-0}\\
Q^{A}_{i}&=&a\partial_{i}\,[f_{A}+\bar{Q}_{A}\upsilon]. \label{eq:Q-A-i}
\end{eqnarray}

In the case of 2 dust fluids, it is natural to choose the total 4-velocity as that of matter: $\upsilon=\upsilon_m$. Then the equations reduce  to
\begin{eqnarray}
\delta_{A}^{\prime}-k^{2}\upsilon_{A}-3\mathcal{H}\Phi^{\prime}&=&\frac{a\bar{Q}_{A}}{\bar{\rho}_{A}}(\Phi-\delta_{A})+\frac{a}{\bar{\rho}_{A}}\,\delta Q_{A}\,,  \label{eq:deltaA-prime}\\
\upsilon_{A}^{\prime} +\mathcal{H}\upsilon_{A}+\Phi&=&\frac{a\bar{Q}_{A}}{\bar{\rho}_{A}}(\upsilon_m-\upsilon_{A})+\frac{af_{A}}{\bar{\rho}_{A}} \,. \label{eq:vA-prime}
\end{eqnarray}
The total matter density contrast evolves according to the equation
\begin{equation}
\delta_{m}^{\prime}-k^{2}\upsilon_{m}-3 \Phi^{\prime} =0 \,. \label{eq:deltam-prime}
\end{equation} 

Our simplified model does not include the small effect of the gravitational potential on the density contrast. Since $|\Phi| \ll |\delta_A|$ on the scales of interest, we can neglect the gravitational potential in \eqref{eq:deltaA-prime} and \eqref{eq:deltam-prime}, so that
\ba
\delta_{A}^{\prime}-k^{2}\upsilon_{A}&=&\frac{a}{\bar{\rho}_{A}}\left(\delta Q_{A}-\delta_{A}\right),  \label{delap} \\
\delta_{m}^{\prime}-k^{2}\upsilon_{m} &=& 0 \,.  \label{delmp}
\ea

Since our model is confined to scales where linear perturbation theory is valid, we can ignore the velocity bias between the halos and the unbound particles. This bias arises on small nonlinear scales because the halos represent peaks in the matter distribution, with a threshold for formation \cite{Biagetti:2014afa,Baldauf:2014fza,Matsubara:2019tyb}. Therefore it is reasonable to assume
\ba
\upsilon_A=\upsilon_m\,.
\ea
Consequently, there is no intrinsic momentum transfer in the matter frame:
\ba
f_A=0\,.
\ea
With these additional simplifications, the perturbed conservation equations reduce to
\ba
\delta_{A}^{\prime}-k^{2}\upsilon_{m}&=&\frac{a}{\bar{\rho}_{A}}\left(\delta Q_{A}-\delta_{A}\right),  \label{delap} \\
\delta_{m}^{\prime}-k^{2}\upsilon_{m} &=& 0 \,,  \label{delmp}\\
\upsilon_{m}^{\prime} +\mathcal{H}\upsilon_{m}+\Phi&=&0 \,. \label{velp}
\ea
By comparing \eqref{eq:deltah-deltam}--\eqref{eq:deltaf1} with \eqref{delap}--\eqref{delmp}, we find that
\ba
\frac{a}{\bar{\rho}_{h}} \delta Q_h &=& \Big[\bar{\mathcal{G}}'+(f-3){\mathcal{H}}\bar{\mathcal{G}} -
3{\mathcal{H}}\bar{\mathcal{G}}^2 \Big]  \delta_m \notag\\
&=& -\frac{a}{\bar{\rho}_{f}} \frac{\Omega_{f}}{\Omega_{h}} \delta Q_f\,.
\ea
Here we used $\delta Q_f=-\delta Q_h$, and the linear growth rate is defined as
\ba
f=\frac{{\rm d}\ln \delta_m}{{\rm d}\ln a} \,,
\ea
so that $ \delta_m'= f \mathcal{H}  \delta_m$.

\begin{figure}[!h]
   \includegraphics[scale = 0.5]{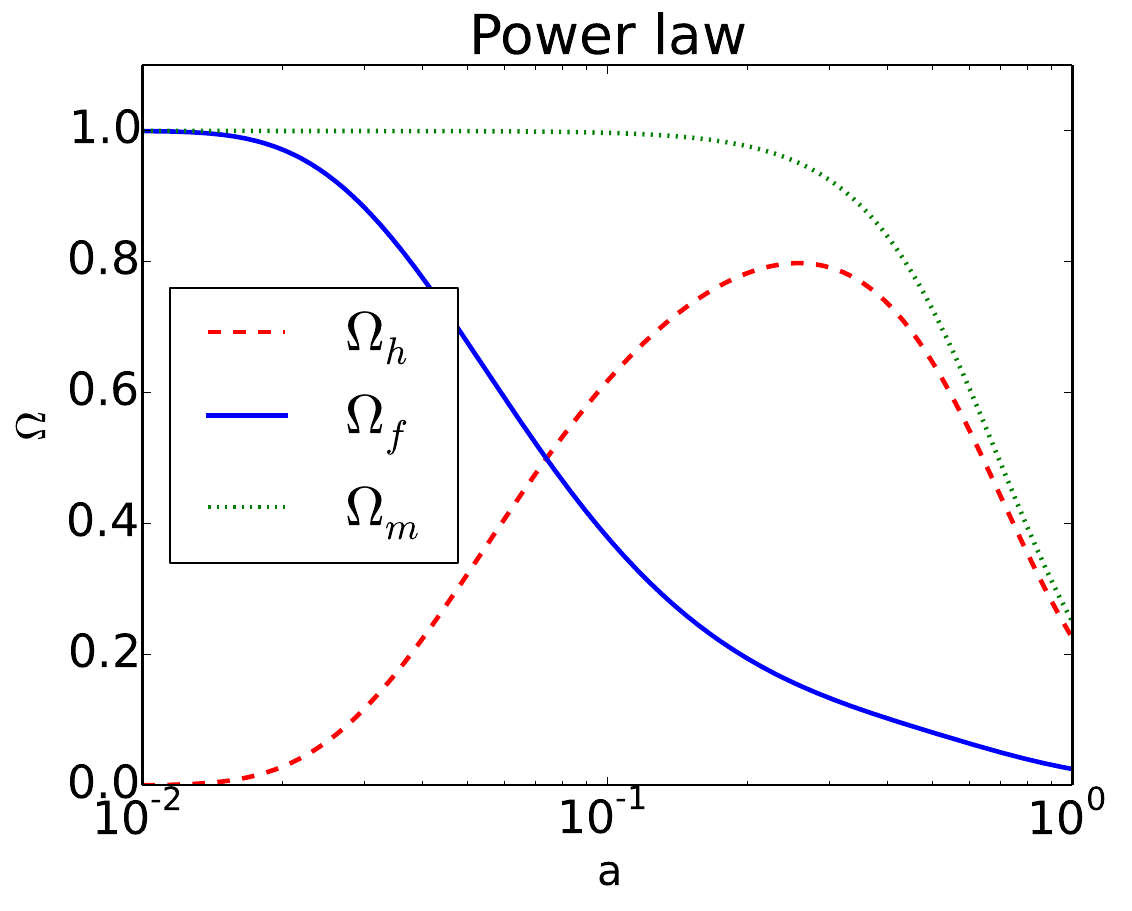}
   \caption{The background density parameters of the bound (halo) fluid $h$ and the unbound (free-particle) fluid $f$, shown together with the total matter  density $\Omega_{m}$, for the model \eqref{pow}--\eqref{eq:domegaf-da}. }
     \label{fig:omegas-power-exponential}
\end{figure}

\section{A simple model}

A simple power-law fit to the results of  \cite{Angulo:2009hf} (their Fig. 3, ellipsoidal collapse model) is
\begin{equation} \label{pow}
\mathcal{F}=\mathcal{F}_{0}\left(\frac{\rho_{m}}{\rho_{m0}}\right)^{-\alpha\left[\left({\rho_{m}}/{\rho_{m0}}\right)^{1/3}-1\right]}\,,
\end{equation}
where the constants are $\mathcal{F}_{0}=0.9$ and $\alpha=0.07$. Note that in the background, $\bar{\mathcal{F}}=\mathcal{F}_{0}(1+z)^{-3\alpha z}$, where $z$ is the cosmological redshift.
The background equations  \eqref{eq:omegah-prime} and~\eqref{eq:omegaf-prime} now become
\begin{eqnarray}
\frac{d\Omega_{h}}{da}&=&\frac{3}{a^{2}}\alpha(1-a-\ln{a})\Omega_{h}+\frac{3}{a}\left(\Omega_{m}-1\right)\Omega_{h} \,,\label{eq:domegah-da}\\
\frac{d\Omega_{f}}{da}&=&-\frac{3}{a^{2}}\alpha(1-a-\ln{a})\Omega_{h}+\frac{3}{a}\left(\Omega_{m}-1\right)\Omega_{f} \,.\label{eq:domegaf-da}
\end{eqnarray}
\begin{figure}
\begin{minipage}[c]{\textwidth}
\centering
\includegraphics[scale=0.42]{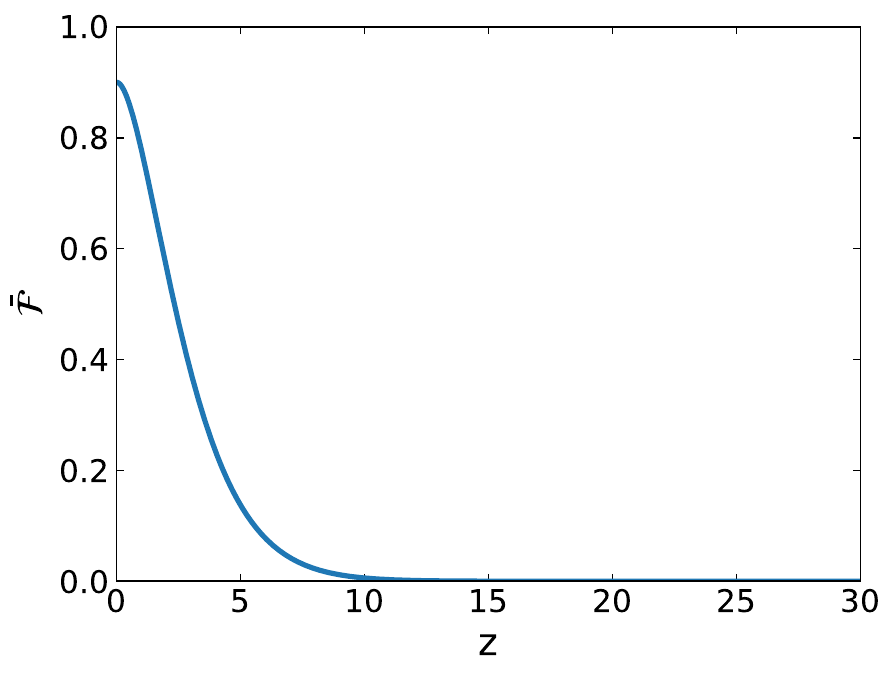}
\includegraphics[scale=0.42]{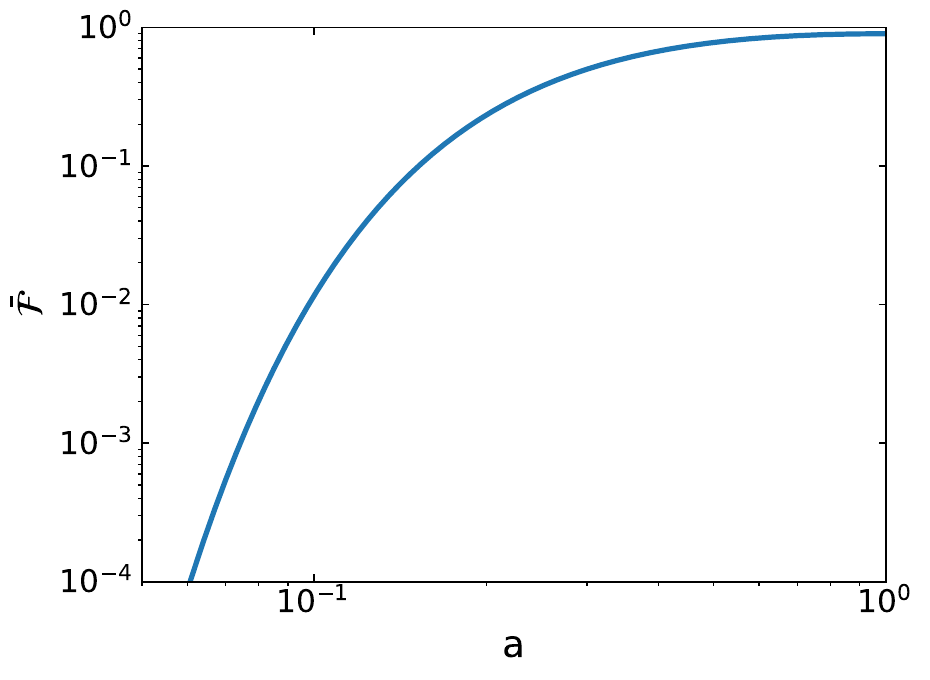}
   \caption{The background density transfer function versus redshift (left) and scale factor (right).
   }
   \label{fig:F-vs-a-z}
   \end{minipage}
\end{figure}

The numerical integration of these equations leads to the background energy density parameters for this model, which are displayed in Figure~\ref{fig:omegas-power-exponential}.
The plot shows how the matter density of the bound structures grows while that of the unclustered matter decreases over cosmic time.  

{The key features of the plot are as follows.}
The high-redshift growth and decay in the matter densities happens until $\Omega_h$-$\Omega_f$ equality is reached at $z\sim 13$ ($a\sim0.07$), when $\Omega_h = \Omega_f \sim0.5$, which is when the clustered matter existed in the same quantity as the unclustered matter. Past this time, $\Omega_h$ continues to grow while $\Omega_f$ continues with its decline.
The halo fluid reaches its highest density ($\Omega_h\sim 0.8$) when $z\sim 5$.
The total matter energy density, $\Omega_m$, remains nearly constant, up to this point, whereafter it starts to decay due to the effect of dark energy,
before reaching its current value of $\Omega_{m0}$.

Plots of the interaction function~\eqref{pow} are displayed in Figure~\ref{fig:F-vs-a-z} where the behaviour of the mass halo fraction with respect to the redshift and the scale factor over cosmic time can be observed.
The fraction of all matter predicted to be in halos at $z=0$, i.e. to be part of clustered matter, is close to hundred per cent. However, at high redshifts or early times a much higher fraction of cold dark matter and cold baryon particles is in diffuse form and is not part of the halos.

 \section{Discussion}
 
We presented a new phenomenological model to describe the evolution of structure formation by treating the matter as composed of two pressure-free components with a purely gravitational interaction: a free-particle fluid (cold dark matter and cold baryons) that is accreted onto the halos, which are the `clustered' particles of the halo fluid.
We derived the background and perturbation equations for the case of two interacting fluids and solved them for the model in which the halo mass fraction $\mathcal{F}$, i.e. the ratio between the bound and unbound matter, is a power law, approximating the results from the  extended Press-Schechter formalism and N-body simulations \cite{Angulo:2009hf}. 
 
The clustered matter grows due to accretion of cold particles, while the unclustered matter density decays over cosmic time. On the other hand, the total matter density, $\Omega_m$, follows the standard evolution, remaining nearly constant for a large part of the cosmic evolution, and decreasing at late times, due to the effect of dark energy.
 
The halo mass fraction, $\mathcal{F}$, grows over cosmic time.
This means that the fraction of all matter predicted to be in diffuse form is much larger at early time while that predicted to be in gravitationally bound structures is much larger at late times, close to hundred per cent at the current epoch.
At low values of $a$, i.e. high redshift, a large fraction of cold particles is free, up to $\Omega_h$-$\Omega_f$ equality at $z\sim13$, beyond which a much higher fraction of cold particles has become part of halos.
These results show that in the framework of hierarchical structure formation, gravitationally bound structures can form not only from the assembly of smaller halos, but directly from free diffuse cold particles (see also \citep{Angulo:2009hf}). 

Our simplistic model is of course no substitute for nonlinear perturbations and simulations, but it does allow for an alternative perspective on some of the general properties of large-scale structure formation. For example, 
Figure \ref{fig:omegas-power-exponential} also reveals the existence of a maximum in halo density at $z\sim 5$. Since galaxies are hosted in halos this feature could be tested against high-redshift galaxy observations, such as those currently underway with JWST (see e.g. \cite{Sabti:2023xwo}).

\[\]
{\bf Acknowledgments:}\\
RM was supported by the South African Radio Astronomy Observatory  and  National Research Foundation (grant 75415). NM acknowledges support by the NRF (grant 111735) and the Anderson-Capelli fund administered by the University of the Witwatersrand DVC office.

\bibliography{Interacting_matter-fluids-epj}

\end{document}